# Fast domain wall motion in nanostripes with out-of-plane fields

Andrew Kunz and Sarah C. Reiff

Physics Department, Marquette University, Milwaukee WI 53233

*Abstract*

Controlling domain wall motion is important due to the impact on the viability of proposed nanowire devices. One hurdle is slow domain wall speed when driven by fields greater than the Walker field, due to nucleation of vortices in the wall. We present simulation results detailing the dynamics of these vortices; including the nucleation and subsequent fast ejection of the vortex core leading to fast domain wall speeds. The ejection is due to the reversal of the core moments by an out-of-plane field. The technique can be used to produce domain walls of known orientation independent of the initial state.

The viability of many proposed devices in magnetic recording, sensing and logic depends on controlling the motion of a domain wall carrying the relevant information [1,2]. In order for the applications to be competitive with present devices, the domain walls must be moved quickly and reliably along thin nanowires. The speed of domain wall driven through a nanowire by an in-plane external magnetic field is maximum when the magnitude of the field is equal to the so-called Walker field [3-6]. When the strength of the field is increased beyond the Walker limit the domain wall begins to move in an oscillatory manner due to the nucleation of (anti-)vortices inside the domain wall [7-9]. These vortices must transit the width of the wire before the domain wall begins to move again. The vortices move slowly across the wire leading to low average speeds. It has been shown that it is possible to inhibit the nucleation process by adding roughness to the wire edges or by artificially removing energy from the precessional motion of the magnetic moments in the field [5, 10-11]. A more recent publication showed that a perpendicularly magnetized underlayer leads to fast wall motion for a large range of driving fields[12]. In this letter, we present simulation results showing that an out-of-plane applied field is responsible for the fast speeds. The field does not inhibit the formation of the anti-vortices, as proposed in reference 12, but instead reversing the polarity of the nucleated anti-vortex core. The gyrotropic force is then responsible for quickly ejecting the vortex from where it was nucleated[13]. Because the anti-vortex is in the wire for only short periods of time the average domain wall speed remains high. The moments in the vortex core are easily controlled and we present a technique for creating a domain wall filter which transmits domain walls of a specific magnetization direction. Domain wall speed, domain wall nucleation and injection, and pinning and releasing from notches all depend on the relative orientation of the magnetization inside the domain wall [6, 7].

In Fig. 1 the average speed as a function of longitudinal in-plane driving field is presented for a domain wall moving through a 100 x 5 nm wire. When no field is applied out of the plane of the wire, the wall behaves as expected; moving quickly below the critical field and slowly above it. When a large out-of-plane field is applied, the average wall speed stays close to the critical speed for all driving fields applied (up to the nucleation field of around 250 Oe for this wire). The inset of Fig. 1 shows the domain wall velocity as a function of out-of-plane field for three wire thicknesses when driven by a 30 Oe in-plane field, a field necessary to induce the oscillatory motion in each wire. The increase in critical out-of-plane field with wire width is due to the reduction of the demagnetizing field. The additional field is necessary to hold the magnetic moments in the plane of the wire necessary inhibiting the anti-vortex formation. A slight dependence on the width of the wire has also been observed.

To understand the behavior of the domain wall it is necessary to review the Landau-Lifshitz (LL) equation of motion for a magnetic moment $m$,

$$\frac{\partial \vec{m}}{dt} = -\gamma \left( \vec{m} \times \vec{H} \right) - \frac{\alpha \gamma}{M_s} \vec{m} \times \left( \vec{m} \times \vec{H} \right) \qquad (1)$$

where $\gamma$ is the gyromagnetic ratio, $\alpha$ is the phenomenological damping parameter, $M_s$ is the magnetization and $H$ is the total field experienced by the moment. The values used for the simulations are consistent with that of permalloy and our simulation procedure has been outlined previously [9]. The right hand side of (1) consists of two primary terms; the first term gives the precessional motion of the magnetic moments and the second is responsible rotating the magnetization into the direction of the local field. A longitudinal in-plane driving field, along the x-axis in Fig. 2, gives the domain wall a velocity along the same axis.

In the presence of a longitudinal in-plane field, the transverse (y-axis) moments of the domain wall are rotated out of the plane of the wire by the first term [14]. The domain snapshot of Fig 2a show the magnetic moments in the domain wall point in the +y direction. The driving field rotates the moments into the $-(\hat{y}\times\hat{x})=+\hat{z}$ direction. When the driving field is greater than the Walker field, an anti-vortex with a core polarization along the z-axis is nucleated at the lower edge of the wire with an initial velocity along the x-axis [14-16]. The vortex core is driven by the gyrotropic force $\vec{F}_g = \vec{G}\times\vec{v}$ where the gyrovector, G, has a direction given by the polarization of the vortex core and $v$ is in the direction of the velocity of the anti-vortex [17,18]. In Fig. 2a the initial gyrotropic force on the nucleated vortex core is in the $\hat{z}\times\hat{x}=+\hat{y}$ direction, pushing the vortex into the wire until it moves out the other side, reversing the direction of the transverse magnetic moments (y-component of the magnetization) in the domain wall as it passes [7,14,19]. In Fig. 2a we plot the transverse position of the vortex core and its polarization direction as a function of time in a wire with zero out-of-plane field and a 30 Oe driving field in a 5 μm long 100 x 5 nm cross-sectional area wire. The domain snapshots represent the direction of the magnetic moments near the domain wall, and the gray-scale shows the z-component of the magnetization. After the transverse moments are reversed the above explanation explains why the vortex core nucleates on the top of the wire and moves downward as time passes in Fig. 2a.

The Zeeman energy term $E_{Zeeman} = -\vec{m}\cdot\vec{H}$ is lowered by aligning the applied field and the magnetic moments. This reduction in energy along with the second term of (1) are responsible for driving the domain wall along the wire; and as shown in Fig. 2b for reversing the direction of the polarization in the anti-vortex core. In Fig. 2b the transverse location (y coordinate) of the anti-vortex core and its polarity is shown as a function of time for a 30 Oe in-plane driving field and an out-of plane field near the critical field. In this case the vortex nucleates and moves into the wire before the out-of- plane field reverses the polarization. Fig. 2b shows that when the polarization direction changes the vortex core reverses direction and the vortex is quickly ejected from the wire by the gyrotropic force. Increasing the magnitude of the out-of-plane field decreases the time necessary to reverse the polarization, which limits the amount of time the vortex exists in the wire. The decrease in time leads to fast domain wall motion.

In Fig. 3 we plot the displacement of a domain wall as a function of time for an initial domain state with the transverse moments in the +y direction and the −y direction under the influence of a large out-of-plane field along with representative domain configurations in a 100 nm wide wire. The gyrotropic force depends on the polarization of the nucleated anti-vortex core. By applying the out-of-plane field in a certain direction, it is possible to create domain walls with a specific transverse direction independent of the starting state. The wall displacements are shifted for ease of viewing. When the transverse wall moments point up (+y) the longitudinal driving field always nucleates an anti-vortex with a polarization in the +z direction due to the first term in (1). An out-of-plane field in the –z direction reverses the moments of the core, ejecting the wall quickly so that the transverse domain wall moments always remain in the +y direction. However, an out-of-plane field in the +z direction pushes the vortex out the opposite side of the wire, reversing the transverse domain wall moments to the –y direction. Similarly, when started in the -y direction, the wall remains down for a positive out-of-plane field and reverses in a negative field. In summary, an out-of-plane field in the positive direction only passes domain walls oriented down; a negative field passes walls oriented up as long as the longitudinal and out-of-plane fields are both greater than their critical field strengths.

Knowledge of the domain wall orientation is important because it has been shown that domain wall velocity depends on the orientation of the moments. Similarly, the field needed to inject a domain wall and the ease of releasing a domain wall from a notch also depends on orientation [6,7].

In conclusion, an out-of-plane field applied externally or with appropriate magnetic under layers can be used to move domain walls at high speed for all driving fields which represents an order of magnitude improvement in useful driving fields. The out-of-plane field has been shown to reverse the magnetic moments nucleated in the anti-vortices which allows for fast removal of the vortex via the gyrotropic force. The direction of the external field can be varied to reliably create domain walls of known orientation independent of their initial orientation [20].

Thanks to Marquette University undergraduate student for creating programs for tracking the anti-vortices and helpful conversations with Oleg Tchernyshyov. This work was supported by the NSF DMR-0706194 and by a Research Corporation Cottrell College Science Award.

**Figure captions**

Fig. 1.  Domain wall speed in a 100 x 5 nm cross-section wire as a function of longitudinal in-plane driving field with and without an additional out-of-plane applied field.  The inset shows the dependence of the critical out-of-plane field on the wire thickness.

Fig. 2.  Plots of the transverse position and polarization of an anti-vortex core without (a) and with (b) an out-of-plane field in a 100 nm wide wire.  The snapshots show the domain configurations with the gray-scale representing the z-component of the magnetization.  In (b) the out-of-plane field reverses the core leading to the fast ejection of the anti-vortex.

Fig. 3.  Wall displacement as a function of time for domain walls of a known initial state for opposite directions of out-of-plane fields in 100nm wide wires.  The cartoons represent the initial and final orientation of the magnetic moments in the domain wall.  A positive out-of-plane field always passes a domain wall oriented down (-y direction).

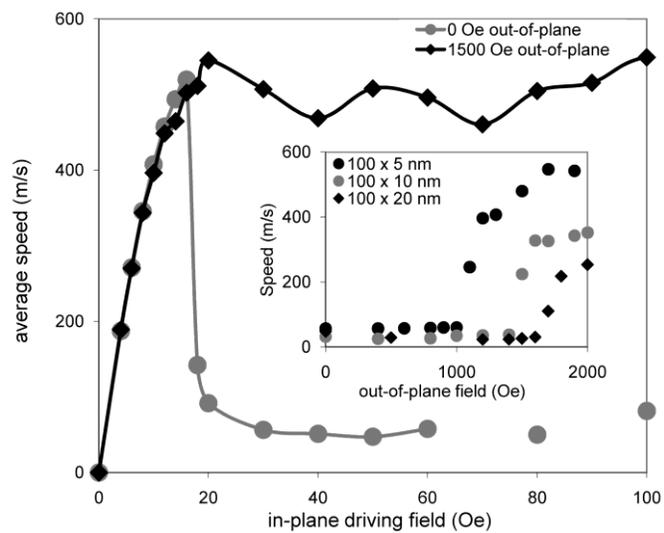

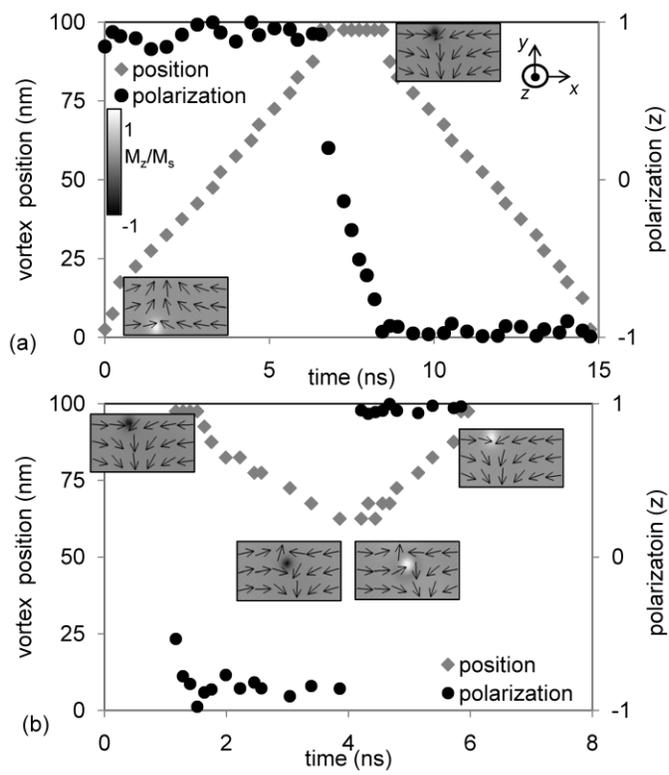

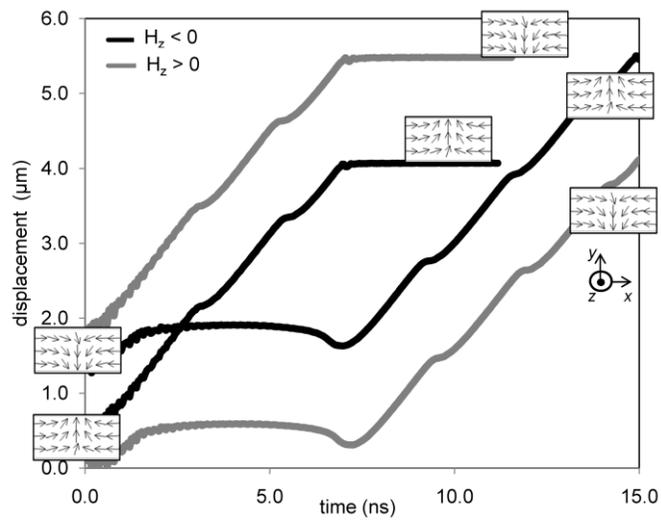

12